\documentclass[a4paper,11pt]{article}
\pdfoutput=1 

\usepackage{jheppub} 

\def\beq{\begin{equation}}

\def\eeq{\end{equation}}

\def\beq{\begin{equation}}

\def\eeq{\end{equation}}

\parskip=\itemsep               

\def\thefootnote{\fnsymbol{footnote}} 

\title{
{\Large  \bf Classicalization and Unitarity.}}
\author{Alex Kovner$^{1}$ and Michael Lublinsky$^{2}$  \\
$^1$ Physics Department, University of Connecticut, 2152 Hillside
Road, Storrs, CT 06269-3046, USA\\
$^2$Physics Department, Ben-Gurion University of the Negev, Beer Sheva 84105, Israel}

\abstract{
We point out that the scenario for UV completion by "classicalization", proposed recently is in fact Wilsonian in the classical Wilsonian sense. It corresponds to the situation when a field theory has a nontrivial UV fixed point governed by a higher dimensional operator. Provided the kinetic term is a relevant operator around this point the theory will flow in the IR to the free scalar theory. Physically, "classicalization", if it can be realized, would correspond to a situation when the fluctuations of the field operator in the UV are smaller than in the IR.
As a result there exists a clear tension between the "classicalization" scenario and constraints imposed by unitarity on a quantum field theory, making the existence of classicalizing unitary theories questionable.
\\
}

\begin{document}
\maketitle
\flushbottom


\voffset1.5cm


\def\thefootnote{\arabic{footnote}} 

\section{Introduction}
Recently Ref.\cite{gia} proposed an interesting scenario for UV completion of theories which are naively perturbatively nonrenormalizable. The mechanism by which this is supposed to happen was dubbed in \cite{gia} "classicalization" and is perceived to be fundamentally different from the classical Wilsonian view of UV renormalizability. We will argue in this note that in fact  classicalization is a variant of the Wilsonian UV completion. The "classicalization" scenario corresponds to the situation where in Wilsonian sense the theory has a nontrivial UV fixed point with relevant operators, which flow to a different fixed point in the IR. In the examples discussed here this IR fixed point is a free massless boson. 
 
The high energy scattering in classicalizing theories is supposed to be dominated by production of multiparticle final states "classical" in nature.
The simplest prototypical model is a theory of a single Goldstone boson with a four derivative interaction with the Lagrangian (although as we point out later, it needs to be modified, if it is to have a chance to be well defined in UV)
\begin{equation}\label{lagr}
L=\frac{1}{2}\partial^\mu\phi\partial_\mu\phi-\frac{g}{\Lambda^4}[\partial^\mu\phi\partial_\mu\phi]^2
\end{equation}
Usually one considers this as an effective field theory below the scale $\Lambda$ obtained by integrating out some massive degrees of freedom with masses of order $\Lambda$. An example would be the "Higgs" field $\rho$ which couples to the  "phase" $\phi$ as
\begin{equation} 
\frac{1}{2}(\rho+\mu)\partial^\mu\phi\partial_\mu\phi
\end{equation}
Integrating out $\rho$ one indeed obtains eq.(\ref{lagr}) with $g=-\lambda^2$, with $\lambda$ being the Higgs self coupling.
\section{Attractive versus repulsive interaction.}
The sign of the coupling $g$ is the determining factor for classicalization. This has been noted already in \cite{gia1}. In our view this is indeed the crucial signature that distinguishes the theories that could potentially classicalize from those that undoubtedly require UV completion in order to be defined in the UV. 

In the case of the negative coupling, $g<0$, 
 the fact that eq.(\ref{lagr}) is only an effective theory and cannot be fundamental reveals itself immediately. Note that for negative $g$ the Euclidean action is not positive definite or bounded from below. The strictly stable configuration is achieved for $\partial_i\phi\rightarrow \infty$. The configuration $\partial_i\phi=0$ is only a local minimum, and thus is unstable. This is of course not a problem in effective theory since only momentum modes of the field $\phi$ with momenta up to $p\sim \Lambda$ are allowed, and the instability is outside the range of the applicability of the model anyway. Nevertheless, this is a clear indication that the cutoff cannot be taken to infinity without modifying the dynamics in the ultraviolet.

On the other hand for positive $g$, the Euclidean action in eq.(\ref{lagr}) is positive definite and does not immediately preclude the possibility that the theory can exist unperturbed all the way to the deep UV. 

Put in different words, let us consider the potential energy for static field configurations.
\begin{equation}\label{pot}
U=\frac{g}{\Lambda^2}[(\partial_i\phi)^2]^2
\end{equation}
For negative $g$ this is attractive interaction. A scalar field with attractive interaction is of course unstable. Even classically such theory collapses with solutions asymptotically tending to $\partial_i\phi\rightarrow\infty$. A new interaction, and possibly new degrees of freedom  must be introduced in the UV to stabilize the system.
For positive $g$ on the other hand, the interaction is repulsive, and there is no obvious inconsistency within the theory itself.

To make this point more explicit, we follow the logic of Ref.\cite{gia}. 
Consider the coupling of the field to a localized static source. The classical equations of motion are
\begin{equation}\label{source}
\partial_i\left[\partial_i\phi+\frac{4g}{\Lambda^4}\partial_i\phi(\partial_j\phi)^2\right]=J\delta(x)
\end{equation}
We concentrate on the properties of a solution, which at large distance, where the field $\phi$ is small, has the perturbative behavior, $\phi(r)\propto J/r$. 

The behavior of the field close to the source depends crucially on the sign of the coupling constant. For negative $g$ the second term in eq.(\ref{source}) counterbalances the first term, and as a result the solution must grow faster than $1/r$ as one approaches the source. In fact one cannot even continue this solution arbitrarily close to the source, as it becomes unstable (oscillating). For positive $g$ on the other hand, the second term in eq.(\ref{source}) has the same sign as the first term, and thus simply takes over some of the burden in reproducing the right hand side. The solution in this case is regular everywhere, and in fact close to the source grows much slower than the perturbative $1/r$. At distance of order $r^2\propto \frac{J{\sqrt g}}{\Lambda^2}$ the two terms in eq.(\ref{source}) become of the same order. For large $J$ this is the typical classical regime, as pointed out in \cite{gia}. At even shorter distances the second term entirely takes over and becomes leading. In this regime the behavior of the field is much softer than in the IR, and the field behaves as $\phi(r\rightarrow 0)\rightarrow (Jr)^{1/3}$, rather than the perturbative $J/r$. The difference between the behaviors of the solutions in the two cases is obviously due to the fact that for $g>0$ the potential is repulsive and thus stable, while for $g<0$ it is attractive and the solution is unstable. 

The situation is the same in other theories, which have been discussed in \cite{gia},\cite{gia1} in connection with classicalization, e.g. the Dirac-Born-Infeld theory
\begin{equation}\label{dbi}
L=\frac{\Lambda^4}{g}\left[1+\frac{g}{\Lambda^4}\partial^\mu\phi\partial_\mu\phi\right]^{1/2}
\end{equation}
 is conjectured to exhibit classicalization for $g>0$, where the interaction of the scalar fields is repulsive. For the attractive interaction, $g<0$ the theory requires a UV completion (see e.g. \cite{greece}).
 \section{Classicalization as a UV fixed point.}
In \cite{gia} the coupling to the source was considered as a proxy of the "seed" that generates the wave function of a "hadronic" state with high energy. Assuming the magnitude of the source to be proportional to the energy, the previous argument suggests that the transverse dimension of the wave function grows as a power of energy, and the "hadron" is a diffuse "classical" object. The scattering cross section of such "black" or "grey" objects is proportional to the area, and thus grows as a power of energy, with the final states dominated by states with large number of particles.

 The situation is reminiscent of QCD at high energies, where in the BFKL approximation scattering probability  initially grows as a power of energy, and unitarization is accompanied by the  growth of the transverse size of hadronic states. Confinement and dynamical mass generation eventually check this growth and bring it back in compliance with the Froissart bound. If not for a finite glueball mass, the cross section in the Yang-Mills theory would indeed grow as a power of energy \cite{urs}. Even with the effects of confinement accounted for, the effective radius of a
 hadron grows with energy logarithmically $r\propto \log (s/m^2)$ due to  the Gribov diffusion.

We note that although the high energy scattering of QCD does conform with the "classicalization" picture advocated in \cite{gia}, the initial rise of the scattering probability in QCD is due to quantum gluon cascade, and not to purely classical behavior of gluon fields. 

Another point this analogy evokes, is that even though the high energy scattering in QCD does probe infrared physics (dominant impact parameters grow as a power of energy, neglecting the finite glueball mass), the theory itself in UV is perfectly Wilsonian and does indeed have an UV fixed point. In a way this is a tautology - any well defined theory has well defined correlation functions at short distances, and behavior of these correlators defines a UV fixed point. The same must be true for the scalar theories discussed in this note as well as in \cite{gia}. If a theory is indeed well defined and does not require introduction of additional degrees of freedom for UV completion, it means that its UV behavior defines a UV fixed point\footnote{In principle more complicated behavior, like a limit cycle is possible. However it is very unlikely in 4 dimensions, as shown in \cite{polchinsky}. We will not consider this possibility in the current note.}.

What type of a UV behavior would characterize theories with classicalization?
Returning to the simple exercise with external source, we can  interpret $J$  as a proxy to the coupling of high momentum modes of the field $\phi$ with its low momentum modes. If the theory is indeed well defined all the way to the UV, one can integrate high momentum modes {\it a la} Wilson to get the Wilsonian effective action eq.(\ref{lagr}). This is true irrespective of the question what type of states dominate high energy scattering, and what is the transverse size of these states. The coupling to an external source produced by the high momentum modes of the field $\phi$, represents the leading contribution to the coupling between the high and low momentum modes of the theory.  Since at short distances the four derivative term dominates the solution, it is the four derivative term that is the more relevant operator in UV. Thus it determines the behavior in the UV, or in other words, determines the nature of the UV fixed point. The kinetic term at this point should be the most IR relevant operator and should determine the IR properties of the theory. Thus a natural interpretation of "classicalizing" theories is that those are theories with nontrivial UV point which in the IR flow to another fixed point: in the case of the model discussed here, a theory of massless free boson.

To roughly understand the properties of the putative UV fixed point, consider the Euclidean action eq.(\ref{lagr}) with $\Lambda$ now meant to stand for the genuine UV cutoff, rather than a cutoff on the effective theory. One expects the Euclidean path integral to be dominated by configurations for which the action is of order unity. This then determines the fluctuations of the field $\phi$  to be of order unity (up to logarithms independent of the cutoff), and so the dynamical dimension of the field $\phi$ at this fixed point is zero, $[\phi]=0$.
The dimensionality of the kinetic term is then (naively) equal to two, indeed confirming that it is a relevant operator. 

This is an unusual situation. If such a UV fixed point exists, the scalar field at this point has very small fluctuations and a much softer behavior that we are normally used to. Even more unusual is the fact that the fluctuations in the UV are softer than the fluctuations in the IR. At the (conjectured) UV fixed point
one expects the correlator to behave as (up to logarithmic corrections)
\begin{equation}
\langle\phi(0)\phi(x)\rangle_{UV}\propto 1
\end{equation}
whereas at the IR fixed point of free massless bosons ($M$ is the UV cutoff of the putative IR free boson theory)
\begin{equation}
\langle\phi(0)\phi(x)\rangle_{IR}\propto \frac{1}{M^2 r^2}
\end{equation}
Usually field fluctuations in the ultraviolet are stronger and thus the correlation function decreases steeply in the UV, with the fall off slowing towards the IR. Naively one expects any theory to have a host of massive excitations in the UV with masses on the order of the cutoff scale. Once the contribution of the heavy states to the correlators have decayed, the rate of change of the correlator should slow down, and thus the fluctuations should decrease. In other words one expects any theory to have more degrees of freedom in the UV than in the IR, and therefore correlations to decrease faster in the UV than in the IR.
\section{Classicalization vs Unitarity.}

Preceding remarks suggest that classicalization scenario may be inconsistent with generic behavior of quantum field theories. In fact, we can identify basic unitarity restrictions on the UV behavior of a scalar theory, which apparently disagree with the idea of classicalization.

 The first potential problem  is a clash with the extension of the Zamolodchikov's "c-theorem" \cite{zamolodchikov} to four dimensions. The four dimensional, so called "a-theorem" was conjectured by Cardy \cite{cardy} and  has been proven recently \cite{zohar}.
The tension between classicalization and the "a-theorem" has been discussed recently in \cite{gia1}. It has been suggested there that the "a-theorem" can be avoided in the classicalizing theories since they do not have a Wilsonian UV completion. From our current point of view however, this argument is highly suspect. Once we identify classicalizing theories as theories with a nontrivial UV fixed point, they become the prime target for the "a-theorem". At this point in fact one cannot definitively tell whether there is a contradiction between the "a-theorem" and the idea of classicalization. There is however a contradiction between the specific model eq.(\ref{lagr}) and its simpe generalizations and the dispersive argument used in the proof of \cite{zohar}. As we argue below this has more general implications for classicalization hypothesis. 

Specifically, ref.\cite{zohar} considers a dilaton Lagrangian of the form eq.(\ref{lagr}) and derives a dispersion relation relating the coupling constant $g$ (assuming $g$ is very small) and the energy integral over the imaginary part of the forward scattering amplitude
\begin{equation}\label{disp}
g\propto -\int_{0^+}^\infty ds \frac{Im \ A(s)}{s^3}
\end{equation}
In a unitary theory $Im A(s)$ has to be positive, since it is proportional to the total scattering cross section. Thus one must have $g<0$ if the dilaton theory is unitary. This conclusion directly contradicts an idea that the theory eq.(\ref{lagr}) classicalizes and is well defined at all scales up to deep UV without the need of introducing additional degrees of freedom\footnote{It has been known for a while that the theory eq.(\ref{lagr}) allows for superluminal propagation around certain backgrounds \cite{nima}. Although we are not sure whether this argument by itself is decisive, we note that the dispersion relation eq.(\ref{disp}) was used in \cite{nima} to illustrate that such theories do not arise at low energy as a result of integration of UV modes.} . 

Some comments are in order here. First off, conforming with the previous argument, the theory eq.(\ref{lagr}) is in fact not unitary and therefore cannot be consistent. This is straightforward to see in the following way.
Even though the four derivative theory eq. (\ref{lagr}) has a positive definite Euclidean action,  its Minkowski space Hamiltonian is not bounded from below even for $g>0$ and therefore the theory has ghosts. To find the Hamiltonian we define, as usual canonical momenta 
\begin{equation}
\Pi(x)=\frac{\delta L}{\delta \partial_0\phi}=\partial_0\phi\left[1-\frac{4g}{\Lambda^4} \partial^\mu\phi\partial_\mu\phi\right]
\end{equation}
To write down the Hamiltonian in terms of canonical variables, one has to express $\partial_0\phi$ in terms of $\Pi$, which in this case involves solving a third order equation. However, we are not interested in the precise canonical structure, but only in the question of boundedness of energy. For this purpose it is sufficient to write the Hamiltonian in terms of $\partial_0\phi$  rather than $\Pi$. This is easily done, and after some trivial algebra we obtain:
\begin{eqnarray}\label{ham1}
H&=&\int d^3x\left[\Pi(x)\partial_0\phi(x)-L\right]=\nonumber \\
&=&\int d^3x\left[\frac{1}{2}(\partial_0\phi)^2+\frac{1}{2}(\partial_i\phi)^2-\frac{g}{\Lambda^4}\left\{(\partial_0\phi)^2-(\partial_i\phi)^2\right\}\left\{3(\partial_0\phi)^2+(\partial_i\phi)^2\right\}\right]
\end{eqnarray}
This Hamiltonian is unbounded from below in the direction $\partial_0\phi\rightarrow \infty$ at fixed $\partial_i\phi$. The instability is reminiscent of theories with ghosts, which are also unbounded from below for large fluctuations of time derivatives of the fundamental fields \cite{ghosts}. The theory defined in eq.(\ref{lagr}) is therefore non-unitary. 

The dispersive argument of ref.\cite{zohar} can indeed be interpreted physically in terms of the ghostlike instability of eq.(\ref{ham1}). Consider a field configuration relevant for a two particle scattering at high energy and low momentum transfer. Such a field clearly has large time derivative, so that $\partial_\mu\phi\partial^\mu\phi\gg 0$. In this kinematics one therefore probes precisely the unstable direction in the phase space of the theory. The usual way such an instability manifests itself in a field theory is via appearance of the ghost states of negative norm. It is thus to be expected that for large enough energies the integrand in eq.(\ref{disp}) becomes negative, since it is dominated by states with negative norm. The dispersion relation eq.(\ref{disp}) thus indeed zeroes in on the non-unitary nature of the theory eq.(\ref{lagr}) with positive coupling \footnote{Note, that, although as discussed above, for negative  coupling the theory eq.(\ref{lagr}) is also unbounded and therefore non-unitary, this does not lead to a contradiction with the dispersion relation. This is due to the fact that the two particle scattering does not probe the direction in the phase space along which the Hamiltonian is unbounded.}.

It may seem strange that a theory with an hermitian Hamiltonian (as eq.(\ref{ham1} presumably is after quantization) exhibits a nonunitary behavior. This however is quite generic if the Hamiltonian is unbounded from below. Some eigenstates of such a Hamiltonian are non normalizable. On such non-normalizable states the Hermiticity property of ordinarily Hermitian operators is lost, and for all intents and purposes the averages behave as if they were calculated in states with negative norm. 
We give an explicit quantum mechanical exampe of this situation in the Appendix. States with negative norm of course violate conservation of probability, thus rendering the theory nonunitary. Schematically, once such (negative norm {\it read} nonnormalizable) states appear as final states in a scattering process (which generically they will), the probability "leaks" through infinity.

In an attempt to rescue the idea of classicalization one can modify the Lagrangian eq.(\ref{lagr}) by adding higher derivative terms, thus rendering energy bounded from below. Consider for example
\begin{equation}\label{lagr2}
L=\frac{1}{2} \partial^\mu\phi\partial_\mu\phi-\frac{g}{\Lambda^4}[ \partial^\mu\phi\partial_\mu\phi]^2+\frac{f}{\Lambda^8}[ \partial^\mu\phi\partial_\mu\phi]^3
\end{equation}
This again has a positive definite Euclidean action for $f>0$. The Hamiltonian is also bounded from below, as is obvious from the following form
\begin{eqnarray}\label{ham2}
H=\int d^3x\left[\frac{1}{2}(\partial_0\phi)^2+\frac{1}{2}(\partial_i\phi)^2-\frac{g}{\Lambda^4}\left\{(\partial_0\phi)^2-(\partial_i\phi)^2\right\}\left\{3(\partial_0\phi)^2+(\partial_i\phi)^2\right\} +\right. \nonumber \\
+\left. \frac{f}{\Lambda^8}\left\{(\partial_0\phi)^2-(\partial_i\phi)^2\right\}^2\left\{5(\partial_0\phi)^2+(\partial_i\phi)^2\right\}\right]
\end{eqnarray}
As we would like the infrared behavior of this theory to be that of a free massless field theory, we should additionally require that the configuration with $\partial_\mu\phi=0$ is the global minimum of energy.
For $g>0$ this is satisfied provided
\begin{equation}
f>\frac{9}{10}g^2
\end{equation}
From the classicalization point of view this theory is not different from the original proposal eq.(\ref{lagr}). The field created by a static source is given by the equation
\begin{equation}
\partial_i\left[\partial_i\phi+\frac{4g}{\Lambda^4}\partial_i\phi(\partial_j\phi)^2+\frac{6f}{\Lambda^8}\partial_i\phi(\partial_j\phi\partial_j\phi)^2\right]=J\delta(x)
\end{equation}
Both nonlinear terms have the same sign as the linear one, and thus the solution with required IR behavior $1/r$ indeed exists. Close to the source the solution has an even softer UV behavior than before
\begin{equation}
\phi(r\rightarrow 0)\rightarrow J^{1/5}r^{3/5}
\end{equation}

The theory defined by the Lagrangian eq.(\ref{lagr2}) is certainly unitary. This however does not ensure that it has a nontrivial continuum limit. For $g,f\ll 1$, the usual perturbative argument tells us that it does not have a nontrivial fixed point. The theory is free, apart from small perturbative corrections, which at energies and momenta much smaller than the cutoff are suppressed as powers of $p^2/\Lambda^2$. In principle it is possible that at some finite (order one) value of (one or both) coupling constants the theory does indeed have a UV fixed point. Similar phenomenon happens in many 3 dimensional theories \cite{fixedpoints}.

However such a situation necessarily means that quantum corrections at the fixed point are large. In particular, even if the "bare" quartic coupling $g$ is positive, it is not assured that the physical renormalized coupling remains so. In fact, the dispersive argument tells us that it must necessarily become negative. In general the dispersive argument  determines the sign of the 1PI four point function, rather than that of the bare coupling. In the context of ref. \cite{zohar} the two are identical, since the dilaton coupling constant is chosen to be very small. In the present context however, this is clearly not the case. If a continuum limit of the theory eq.(\ref{lagr2}) exists, it is of course unitary, since the original UV Hamiltonian is bounded from below. The scattering amplitude $A(s)$ then cannot grow as fast as $s^2$ for large $s$, and the integral on the right hand side of eq.(\ref{disp}) therefore is convergent and positive. We are thus forced to conclude that the physical quartic coupling in the infrared is negative. 

How does this reflect on the classicalization idea? Our discussion (as well as that of ref.\cite{gia}) of eq. (\ref{source}) has been very schematic. However, if one really understands this equation as detemining the structure of low momentum field in high energy states, the only sensible way to understand the field $\phi$ that enters this equation is as the expectation value of the field operator in the presence of the external source. The coefficient $g$ therefore should be understood as the 1PI Green's function. In this equation therefore $g$ must be negative if we are trying to determine the structure of the field far away from the source, while, as discussed above, classicalization requires positive coupling $g$\footnote{Negative $g$ would lead to larger field fluctuations in the UV, and thus would not violate the unitarity bound of \cite{mack}.}.
We therefore have to conclude that the dispersive argument (or equivalently unitarity of the theory) is in contradiction with the classicalization scenario.

It is now also clear that although our discussion has been shaped after the theory eq.(\ref{lagr}), it in fact applies to many thoeries of this type. For example, the "classicalizing" version of the Dirac-Born-Infeld theory eq.(\ref{dbi}) is afflicted with the same malady. It requires a positive quartic coupling as is seen from the expansion of eq.(\ref{dbi}) in powers of $(\partial_\mu\phi)^2$ and thus is the subject of the above dispersive argument. In fact, the DBI Lagrangian is even more suspect than eq.(\ref{lagr}), as it is not a real function\footnote{An attempt to make sense of theories of this type has been made in \cite{greece}. It is however not at all clear that the approach of \cite{greece} can yield a well defined quantum field theory.}.

 We can sharpen the problem by considering the general unitarity bound on possible values of dimension of a scalar field in any conformal theory derived in \cite{mack}. It states that if a four dimensional conformal quantum field theory is unitary, dimensions of primary scalar fields in it must satisfy $[\phi]\ge 1$, with equality holding only for free noninteracting scalar. Our dimensional analysis of putative UV fixed points strongly suggests, that classicalization must lead to the propagator of  the field $\phi$ which decreases with distance slower in the UV than in the IR. This is tantamount to $[\phi]<1$ at the ultraviolet fixed point. The unitarity bound of \cite{mack} prohibits such a behavior\footnote{We note that there is no similar argument for theories of gravity. There does exist a lower bound on the dynamical dimension of a tensor field, $d\ge 4$. This bound however only applies to gauge invariant tensors and thus does not constrain the fluctuations of the metric. In fact, since by definition, no tensor in a theory with general coordinate invariance is gauge invariant, this bound is meaningless. The bound on the gauge invariant scalar $R$ still exists, $d\ge 1$. However since in the linearized gravity the dimension of $R$ is equal to four, this bound is not in contradiction with considerable softening of the fluctuations of the gravitational field in the UV \cite{DG,mazumdar}. Even here though, one has to be mindful that the allowed suppression is powerlike, and not exponential as in some of the suggestions in \cite{mazumdar}.} which may have softer UV behavior . This argument does not depend on a specific form of the Lagrangian, but merely on an assertion that in the IR the theory describes free boson particles. Admittedly, one would require an explicit calculation of the UV propagator in order to make this argument more quantitative.

In conclusion, we have presented arguments to the effect that the UV behavior necessary for classicalization cannot be realized in unitary Quantum Field Theories. It would be very valuable to put these arguments on a quantitative basis, or alternatively to understand how they are avoided if classicalization is to remain feasible.

\section{Appendix}
In this appendix we illustrate how nonnormalizable states can camouflage as states with negative norm.
Consider an utterly trivial example of a harmonic oscillator with negative Hamiltonian
\begin{equation}\label{hamhar}
H=-\frac{\pi^2}{2}-\frac{\omega^2}{2}x^2
\end{equation}

Let us define the usual creation and annihilation operators
\begin{equation}
a=\sqrt{\frac{\omega}{2}}x+i\sqrt{\frac{1}{2\omega}}\pi_x
\end{equation}
The Fock vacuum of $a$ is the normalized Gaussian state
\begin{equation}
a|0\rangle=0; \ \ \ \ \ |0\rangle=e^{-\frac{\omega}{2}x^2}
\end{equation}
This is the state with highest energy. The vacuum of $a^\dagger$ is 
\begin{equation}
a^\dagger|\Phi\rangle=0; \ \ \ \ \ |\Phi\rangle=e^{\frac{\omega}{2}x^2}
\end{equation}
This state is nonnormalizable and in general quite nonsensical, but formally corresponds to the lowest energy state of the Hamiltonian eq.(\ref{hamhar}). If we are not careful, the standard formal manipulations in quantum mechanics can push us to work close to this state. For example choosing the standard $i\epsilon$ prescription in a path integral projects the theory onto this state. 

The Hamiltonian can be written as
\begin{equation}
H=-\omega aa^\dagger + E_0
\end{equation}
Let us consider a tower of states above $|\Phi\rangle$ generated by the action of operators $a$.
We then have
\begin{equation}
H|\Phi\rangle=E_0|\Phi\rangle; \ \ \ H|1\rangle\equiv Ha|\Phi\rangle=E_0a|\Phi\rangle-\omega a a^\dagger a|\Phi\rangle=(E_0+\omega)|1\rangle
\end{equation}
Thus applying operator $a$ increases the energy of the state $\Phi\rangle$ by $\omega$, and the spectrum seems to be bounded from below. However if we blindly assume that these states satisfy the usual properties of the vectors in the Hilbert space, we quickly run into a paradox, namely that some of them have negative norm. Consider the following formal argument. Let us calculate the norm of the "`one particle state"'
\begin{equation}
\langle1|1\rangle=\langle\Phi|a^\dagger a|\Phi\rangle=\langle\Phi|aa^\dagger-1|\Phi\rangle=-\langle \Phi|\Phi\rangle
\end{equation}
We may (rushly) conclude from this argument that either the one particle state or the vacuum state must have negative norm. This is of course incorrect. The "vacuum" state is simply a divergent Gaussian, and has a divergent rather than negative norm. The one particle wave function we can find explicitly
\begin{equation}
a|\Phi\rangle=\left(\sqrt{\frac{\omega}{2}}x+\sqrt{\frac{1}{2\omega}}\frac{d}{dx}\right)e^{\frac{\omega}{2}x^2}=\sqrt{2\omega}xe^{\frac{\omega}{2}x^2}
\end{equation}
The norm of this state is also divergent and in no way negative. The flaw in the argument is of course precisely the fact that the states we are dealing with are not normalizable and therefore integration by parts is not allowed. In particular
\begin{equation}
\langle\Phi|\Phi\rangle\ne |a|\Phi\rangle|^2
\end{equation} 
as is trivially verified by an explicit calculation. In fact the difference between the two sides of the inequality is infinite. 

For any of these nonnormalizable states the probability easily "`leaks"' through infinity, since most of the probability is localized infinitely far from the origin in the first place.
In this "`quantization scheme"' therefore the unitarity is broken.  One can introduce an infrared regulator which makes the norm finite, but the wave functions will still peak at the regulator scale, rather than inside the bulk of space. In a noninteracting theory this is not a problem, since an initial state which is localized close to the origin never mixes through evolution with any of these nonnormalizable states. However if this ghost oscillator is nontrivially coupled to a positive energy mode, the evolution can easily excite the nonnormalizable states since unbounded amounts of energy can flow from the ghost sector to the "`matter"' sector, see for example \cite{trodden}.  Such states then run great risk of leaking the probability through the boundary under time evolution.  This happens even if the full Hamiltonain naively looks perfectly Hermitian.

{\it Note Added}

After this paper was submitted, the preprint \cite{vikman} appeared, which using a different approach also concludes that fluctuations in UV in classicalizing theories are suppressed.

\section*{Acknowledgments}
The work of AK is supported by DOE grant DE-FG02-92ER40716.
The work of ML is partially supported by the Marie Curie Grant  PIRG-GA-2009-256313. AK thanks the Physics Department of Ben-Gurion University of the Negev for hospitality during the visit when this work has been initiated.


\end{document}